\documentclass[11pt]{article}

\usepackage{graphicx,apalike}

\newcommand{\etal}{\textit{et al.}}

\newcommand{\ie}{i.e.}

\newcommand{\given}{\! \mid \!}
\newcommand{\phihat}{\hat{\phi}}
\newcommand{\Phihat}{\hat{\Phi}}
\newcommand\psihat{{\hat{\psi}}}
\newcommand\thetahat{{\hat{\theta}}}
\newcommand{\kappahat}{\hat{\kappa}}
\newcommand{\boldxi}{\mathbf{A}}

\newcommand{\eqref}{\ref}

\newcommand{\doubleint}{\int\!\!\!\int}
\newcommand\tripleint{{\int\!\!\!\int\!\!\!\int}}

\newcommand{\set}[1]{\ensuremath{ \left \{#1 \right \}}}
\newcommand{\spanset}[1]{\ensuremath{\mathrm{span}\set{#1}}}

\newcommand{\Ptilde}{\tilde{P}}

\begin{document}

\title{{\LARGE \bf Neural Representation of Probabilistic Information}}
\author{\bf 
    M.J.~Barber\thanks{Institut f\"ur Theoretische Physik, 
    Universit\"at zu K\"oln,
    D-50937 K\"oln,
    Germany} \and 
    \bf J.W.~Clark\thanks{ Department of 
Physics, Washington University, Saint Louis, MO 63130}    \and
    \bf C.H.~Anderson\thanks{Department of Anatomy and Neurobiology, Washington University 
School of Medicine,  Saint Louis, MO 63110}
}

\date{\today}    
\maketitle

\abstract{It has been proposed that  populations of neurons process 
information in terms
of probability density functions (PDFs) of analog variables.  
Such analog variables range, for example, from target luminance 
and depth on the sensory interface to eye position and joint
angles on the motor output side.  The requirement that
analog variables must be processed leads inevitably to
a probabilistic description, while the limited precision
and lifetime of the neuronal processing units leads naturally
to a population representation of information.  
We show how
a time-dependent probability density $\rho(x;t)$ over variable
$x$, residing in a specified function space of dimension $D$, 
may be decoded from the neuronal activities in a population as 
a linear combination of certain decoding functions $\phi_i(x)$, 
with coefficients given by the $N$ firing rates $a_i(t)$ (generally
with $D << N$).  We show how the neuronal encoding process may
be described by projecting a set of complementary encoding functions 
${\hat \phi}_i(x)$ on the probability density $\rho(x;t)$, and 
passing the result through a rectifying nonlinear activation 
function.  We show how both encoders $\hat \phi_i(x)$ and decoders
$\phi_i(x)$ may be determined by minimizing cost functions that 
quantify the inaccuracy of the representation.  Expressing a given 
computation in terms of manipulation and transformation of probabilities, 
we show how this representation leads to a neural circuit that can 
carry out the required computation within a consistent Bayesian 
framework, with  the synaptic weights being explicitly generated
in terms of encoders, decoders, conditional probabilities, and priors.
}
\break

\section{Introduction}

It has been hypothesized (Anderson, 1994, 1996)
\nocite{anderson:1994,anderson:1996} that circuits of 
cortical neurons perform statistical inference,
and, in particular, that they encode and process information about analog 
variables in the form of probability density functions (PDFs).  This 
PDF hypothesis provides a unified framework for understanding 
diverse observations from experimental neurobiology, constructing 
neural network models, and gaining insights into how neurons can 
implement a rich collection of information-processing functions.

The PDF hypothesis derives from two major themes of computational 
neuroscience.  The first theme stems 
from efforts to determine how information is represented by neural 
systems, through understanding how neural activity correlates to external 
cues or actions (such as sensory stimuli or motor response).  Our 
understanding of neural encoding can be tested by inferring sensory 
input or motor output from a set of neural activities, and comparing 
 the estimate thus obtained to the external cue or action.

To decode the response from a population of neurons requires 
procedures to infer information from individual spike trains, as well 
as procedures to combine these results into an aggregate estimate.  
An optimal method for decoding information from individual neural spike trains 
has been developed \cite{bialek/etal:1991,bialek/rieke:1992,rieke/etal:1997} and applied to 
movement-sensitive neurons in the blowfly \cite{rieke/etal:1997} 
and to other systems 
\cite{theunissen/etal:1996}.  This method consists of utilizing a linear filter to 
extract the maximum possible information from each spike (typically a 
few bits; see Rieke \etal, 1997\nocite{rieke/etal:1997}), 
as measured by the ability to reconstruct the 
stimulus from the spike train.  In these studies, the linear filter 
determines a firing rate from the spike trains; this firing rate 
contains most of the information, with additional information possibly 
encoded in other aspects of the activity patterns.  In the current work, 
we assume that the 
firing rates capture the essential behavior of neural systems, and 
will not explicitly consider spike trains.

Methods for decoding information from the firing rates of 
populations of neurons were pioneered by Georgopoulos and 
collaborators.  They showed that a ``population vector'' derived from 
the firing rates of a population of cortical neurons can be used to 
predict the intended arm movements of monkeys 
\cite{georgopoulos/etal:1986,schwartz:1993}.  This vector estimate of direction, 
$\mathbf{V}_\mathrm{est}$, is obtained from the neural firing rates 
$a_{i}$ by
\begin{equation}
    \mathbf{V}_\mathrm{est} = \sum_{i=1}^{N}a_{i}\mathbf{C}_{i}
    \label{eq:intropopvec}
\end{equation}
where the preferred direction vectors, $\mathbf{C}_{i}$, indicate the 
direction at which neuron $i$ has its maximal firing response.  The 
population vector approach has been refined and extended by several 
authors; in particular, Salinas and Abbott 
(1994\nocite{salinas/abbott:1994}) provide an excellent 
discussion of several such refinements, as well as introducing their 
own.  The emphasis in such studies has been the reconstruction of 
vector quantities from populations of neural responses by a process 
that in several cases appears to be computation of an expectation 
value from an implicit probability distribution.

The second theme leading to the PDF hypothesis stems from an analysis 
showing that the original Hopfield neural network implements, in 
effect, Bayesian inference on analog quantities in terms of PDFs 
\cite{anderson/abrahams:1987}.   The role of PDFs in neural 
information processing 
is  being
explored along a number of avenues.  As in the present work,
Zemel \etal\ (1998\nocite{zemel/etal:1998}) have
investigated population coding of probability distributions, but with 
different representations than those we will consider here.  Several 
extensions of this representation scheme have been developed 
\cite{zemel:1999,zemel/dayan:1999,yang/zemel:2000} that 
feature information propagation between interacting neural populations.  
Further, a number 
of related models have been introduced.  Of particular note is a dynamic routing 
model of directed attention (Anderson and Van Essen, 1987; Olshausen 
\etal, 
1993, 
1995\nocite{anderson/vanessen:1987,olshausen/etal:1993,olshausen/etal:1995}).  
Additionally, several ``stochastic 
machines'' \cite{haykin:1999} have been formulated, including Boltzmann 
machines \cite{hinton/sejnowski:1986}, sigmoid belief networks 
\cite{neal:1992}, and Helmholtz machines \cite{dayan/hinton:1996}.  Stochastic 
machines are built of stochastic neurons that choose  one of two 
possible states in a probabilistic manner. Learning rules for 
stochastic machines enable such systems to model the underlying probability 
distribution of a given data set; however, they are not biologically 
realistic.  

The 
two prominent themes of population coding and probabilistic inference
are combined in the PDF hypothesis through the assertion 
that a physical variable $x$ is described 
by a neural population at time $t$ in terms of a PDF 
$\rho(x;t)$, rather than as a single-valued estimate $x(t)$.  Such a PDF 
description has the significant advantage that it not only permits a 
single-valued estimate to be calculated, but also provides for  
measures of the uncertainty of such estimates.  For example, a 
specific value $\xi$ at time $t$ can be represented as the mean of a 
normal distribution over $x$ with variance $\sigma^{2}$, so that
\begin{equation}
    \rho(x;t) = N\left(x; \xi(t), \sigma^{2}(t)\right)
\end{equation}
Clearly, this PDF allows $\xi(t)$ to be known very precisely (small 
variance) or with a great deal of uncertainty (large variance).

More generally, we consider a PDF described at time $t$ in terms of a 
set of $D$ underlying parameters \set{A_{\mu}}. 
Guided by the experimentally observed linear decoding rules discussed above, 
we will take the PDFs to be represented by
\begin{equation}
    \rho(x;t) \equiv \rho\left(x; \set{A_{\mu}(t)} 
    \right) = \sum_{\mu=1}^{D}A_{\mu}(t) \Phi_{\mu}(x)
    \label{eq:basisdec}
\end{equation}
The basis functions $\Phi_{\mu}(x)$ are orthonormal functions that define 
the PDFs  the neural circuit can represent.
We describe 
$x$ with $\rho(x; \set{A_{\mu}(t)})$ rather than 
$\rho(x \given \set{A_{\mu}(t)})$ to distinguish 
between the assumed forms of models (equation~%
\ref{eq:basisdec})
and 
relationships that exist amongst random variables (viz.\ 
conditional probabilities).

The amplitudes $A_{\mu}(t)$ of the representations defined by 
equation~%
\ref{eq:basisdec} cannot be interpreted as neuronal firing 
rates: they can take on negative values and are more precise 
than neuronal firing rates.  However, we can represent a PDF in terms of 
decoding functions $\phi_{i}(x)$ and firing rates $a_{i}(t)$ 
associated with $N$ neurons, so that
\begin{equation}
    \rho(x;t) = \sum_{i=1}^{N}a_{i}(t)\phi_{i}(x)
    \label{eq:decoderequation}
	\label{eq:xencoderule}	
\end{equation}
Unlike the basis functions $\Phi_{\mu}(x)$, the decoding functions 
$\phi_{i}(x)$ form a highly redundant, overcomplete representation 
($N \gg D$) that 
is specialized for use with neurons of limited precision.

From the relations asserted in equations~\ref{eq:basisdec} 
and~\ref{eq:decoderequation}, we can identify three relevant problem
domains.  First, we have the physical variable~\(x\), 
described by the PDF~\(\rho(x;t)\).  This domain is that of
high-level concepts.
Second, we have the neural
network with its measurable neural firing rates~\( a_i(t)\).  
The neural
network constitutes a physical implementation of the desired
computations on the physical variable, so the properties of this second
domain should be chosen to match the properties of biological
systems as closely as possible.  In particular, the neural firing
rates must be constrained to be positive quantities of low precision.  The
third domain is that of the underlying parameters~\(A_{\mu}\), which
subserve an alternative, abstract implementation of the desired computations.
The constraint in this case is
minimality:  
we concern ourselves
only with mathematical convenience and allow the~\(A_{\mu}\) to be 
of arbitrary precision and to take on negative values.

Following Zemel \etal\ (1998\nocite{zemel/etal:1998}),  the 
domain of physical 
variables is called the \emph{implicit space} and the domain of 
measurable quantities 
the \emph{explicit space}.  Extending their nomenclature, we  
shall refer to the third domain as the \emph{minimal space}.  The minimal
space will serve as a useful bridge between the two other spaces.

It may be conceptually helpful to regard the variables or parameters
$A_\mu(t)$ as the activities of a set of $D$ ``metaneurons,'' fictitious
entities that reside and act in the minimal space.  However, it must 
be emphasized that such metaneurons differ from real neurons in their abilities 
to function with high precision and to produce negative ``firing rates''
$A_\mu(t)$.  Accordingly, they possess valuable properties that
will facilitate formal representation and analysis.

\section{Obtaining the Neuronal Representation}\label{sec:encodechap}

\subsection{Multiple Levels of Representation}

The fundamental assumption of the framework to be developed in this
paper is that information about a 
physical variable~$x$ given a set of parameters $\mathbf{A} = 
\set{A_{1},A_{2},A_{3},\ldots}$ at time~$t$ is represented by an 
ensemble of neurons as a PDF $\rho(x; 
A_{1}(t), A_{2}(t), A_{3}(t),\ldots)$.  For notational 
convenience, we will usually abbreviate this quantity as 
$\rho(x;t)$.  This PDF can be determined from a set of neuronal firing 
rates $\{a_i(t)\}$ using a set of decoding functions (or simply 
decoders) $\phi_i(x)$, as prescribed in equation~\ref{eq:decoderequation}.
In turn, a set of encoding functions (encoders) $\phihat_i(x)$ is used to determine the 
firing rates from an assumed PDF by means of
\begin{equation}
	a_i(t) = f \left (\int \phihat_i(x) \rho(x;t) dx \right )
	\label{eq:encoderequation}
\end{equation}
where a nonlinear activation function~$f()$ is introduced to preclude negative 
firing rates.  The encoding functions \( \phihat_{i}(x) \) must be 
chosen so as to yield a close match to desired (\ie\ experimentally 
observed) firing rates \( a_{i}(t) \).
The decoding rule 
(equation~\ref{eq:decoderequation}) should in general be viewed as only returning 
an approximation to the PDF: in particular, functions that are not strictly positive 
semidefinite can be decoded from such a rule.

We can also represent the PDF using a complete 
orthonormal basis \set{\Phi_{\mu}(x)} for the space spanned by the 
decoders, as shown in equation~\ref{eq:basisdec}.
Further, we can represent the decoding functions in terms of this 
basis, writing
\begin{equation}
	\phi_{i}(x) = \sum_{\nu=1}^{D} \kappa_{\nu i}\Phi_{\nu}(x)
	\label{eq:Phitophi}
\end{equation}
where the $\kappa_{\nu i}$ are coupling coefficients to be determined.  
Since we now have an orthonormal basis, the coefficients $A_{\mu}$ in 
equation~\ref{eq:basisdec} are simply evaluated from
\begin{equation}
A_{\mu}(t) = \int \Phi_{\mu}(x) \rho(x ;t) dx 
\label{eq:minspaceencoderule}
\end{equation}
The encoding and decoding rules based on the amplitudes~\( A_{\mu}(t) 
\) in the minimal space are seen to parallel those based on the 
neuronal firing rates \( a_{i}(t) \), apart from the absence of a 
nonlinearity in equation~\ref{eq:minspaceencoderule}.
In this section, we will 
develop methods to relate operations in the mathematically convenient minimal space 
and the biologically plausible implementation of PDFs in the 
explicit space of model neurons.

\subsection{Obtaining the Encoding Functions}\label{sec:learnencoders}

Although we do not know the encoding functions at 
this point, we do know that they can be represented in terms of another 
set of basis functions \set{\Phihat_{\mu}(x)} through
\begin{equation}
	\phihat_{j}(x) = \sum_{\mu}\kappahat_{j\mu}\Phihat_{\mu}(x)
\end{equation}
where the coupling coefficients $\kappahat_{j\mu}$ are in general 
distinct from the $\kappa_{\nu i}$.
For many networks, it is appropriate to assume the basis for the encoders to 
be identical to the basis for the decoders.  For example, in the case 
of the neural integrator (see section~\ref{sec:neuralintegrator}) the PDFs 
are continually mapped into and out of the minimal space provided by 
the $\Phi_{\mu}(x)$ and 
$\Phihat_{\nu}(x)$.  Thus,   
\spanset {\Phi_{\mu}} can be equal to \spanset {\Phihat_{\nu}}.  
For definiteness, we take $\Phi_{\mu}(x) = \Phihat_{\mu}(x)$.

To find the encoding functions, we define the cost function
\begin{eqnarray}
	E_{1} &=& {1 \over 2} \sum_{i} \int \left [ a_{i}(\boldxi) - f \left (
\int \phihat_{i}(x) \rho(x; \boldxi) dx \right ) \right 
]^{2}\rho(\boldxi) d\boldxi \nonumber \\
&=& {1 \over 2} \sum_{i} \int \left [ a_{i}(\boldxi) - f \left 
(\sum_{\nu} \kappahat_{i\nu} \int \Phi_{\nu}(x) \rho(x; \boldxi) dx 
\right ) \right ]^{2}\rho(\boldxi) d\boldxi \nonumber \\
&&
\end{eqnarray}
We now use gradient descent to determine the $\kappahat_{i\nu}$ that 
minimize $E_{1}$
\begin{equation}
	{d\kappahat_{j\mu} \over dt} \approx -\eta{ \partial E_{1} \over 
\partial \kappahat_{j\mu}} = \int \left(  a_{j}(\boldxi) - f ( 
h_{j}(\boldxi) ) \right ) f' (h_{j}(\boldxi)) U_{\mu}(\boldxi) 
\rho(\boldxi) d\boldxi
\end{equation}
where \( \eta \) is a rate constant.  We have defined
\begin{eqnarray}
U_{\nu}(\boldxi) = \int \Phi_{\nu}(x) \rho(x; \boldxi)dx \\
h_{j} (\boldxi) = \sum_{\nu}\kappahat_{j\nu} U_{\nu}(\boldxi)
\end{eqnarray}
to simplify the expression.

\begin{figure}[tbp]
    \centering
    \includegraphics{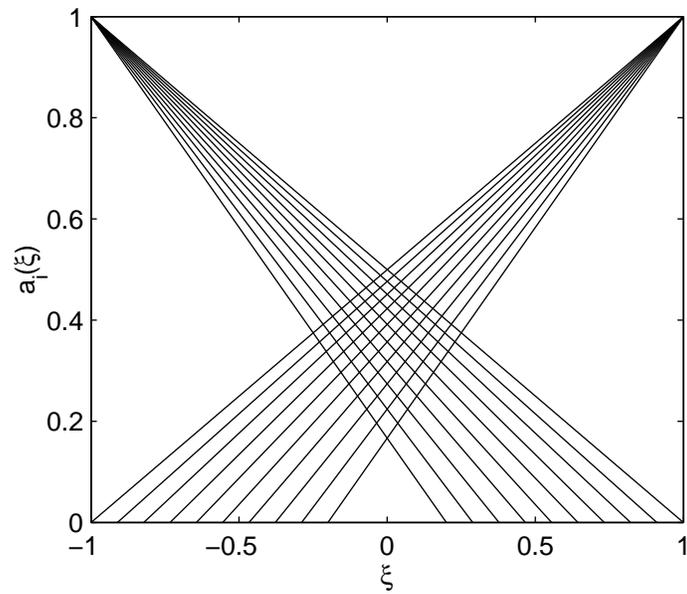} 
    \caption{Broad
    piecewise-linear functions provide biologically plausible neural 
    firing rates.  These firing profiles are similar to one-dimensional 
    versions of the neural responses used to construct
    Georgopoulos's population 
    vector.}
    \label{fig:broadroughact}
\end{figure}

\begin{figure}[tbp]
    \begin{center}
	\begin{tabular}{l}
	    a  \\
	    \includegraphics[width=2.25in]{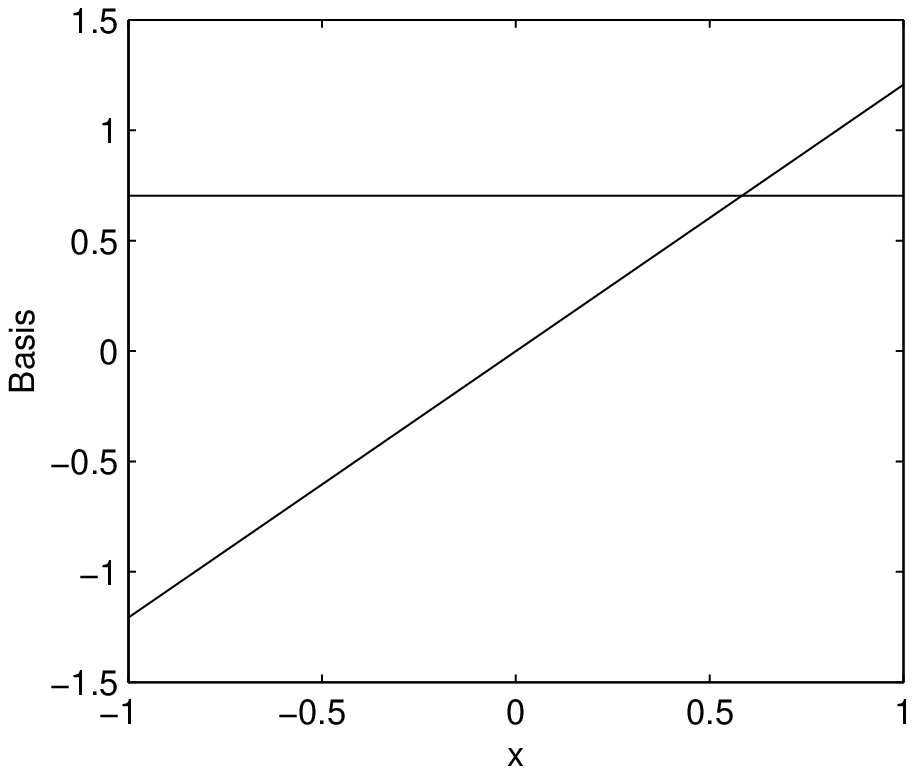}  \\
	\end{tabular}
	\hfill
	\begin{tabular}{l}
	    b  \\
	    \includegraphics[width=2.25in]{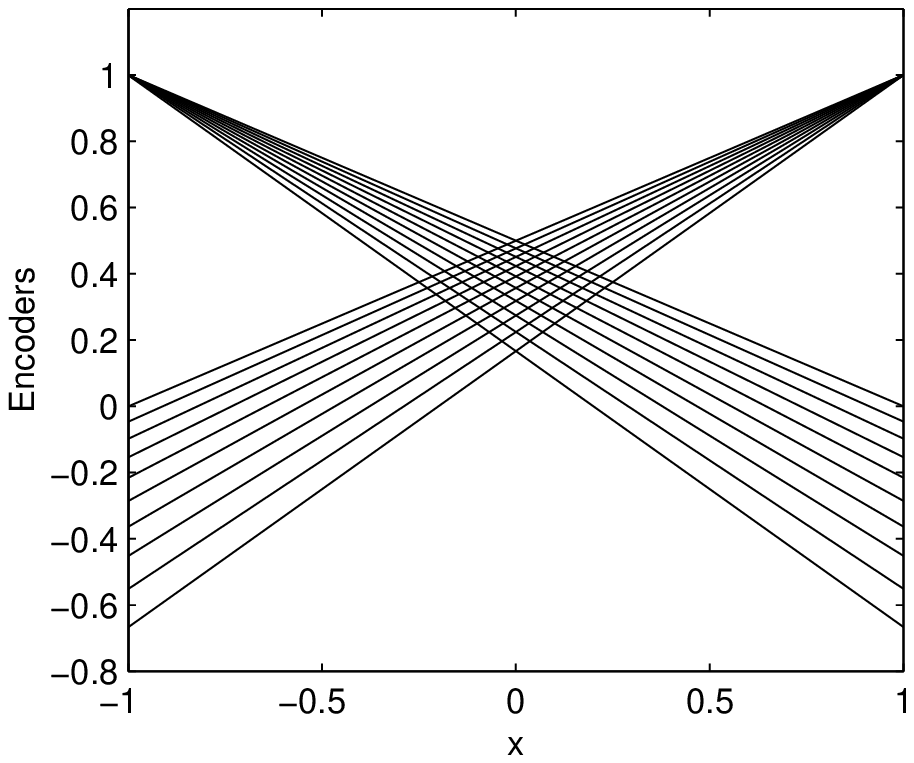}  \\
	\end{tabular}
	\begin{tabular}{l}
	    c  \\
	    \includegraphics[width=2.25in]{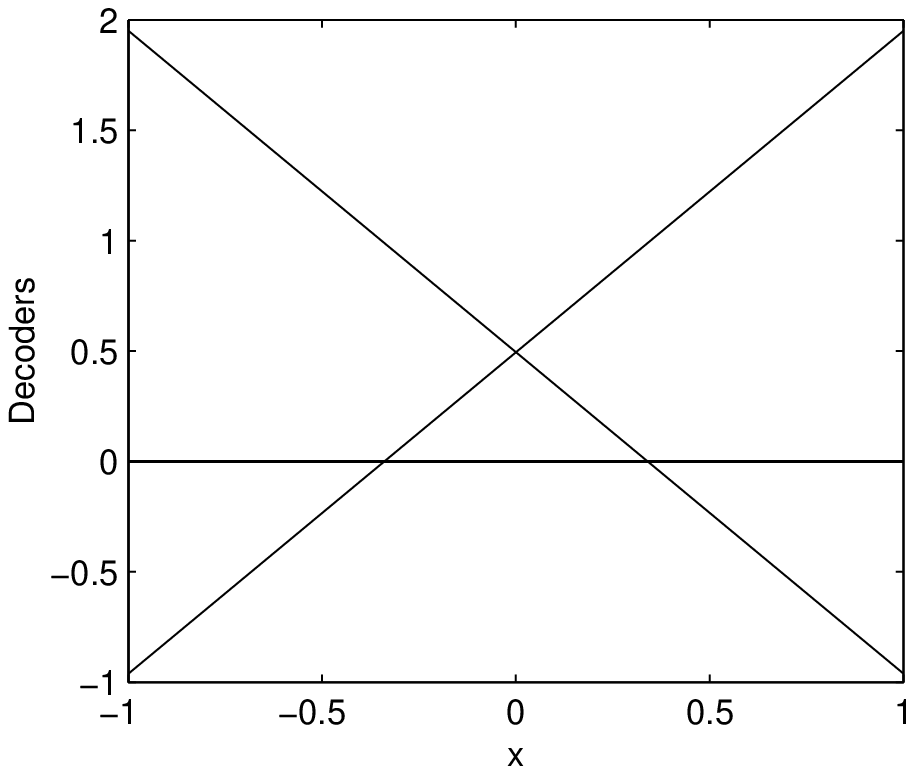}  \\
	\end{tabular}
	\hfill
	\begin{tabular}{l}
	    d  \\
	    \includegraphics[width=2.25in]{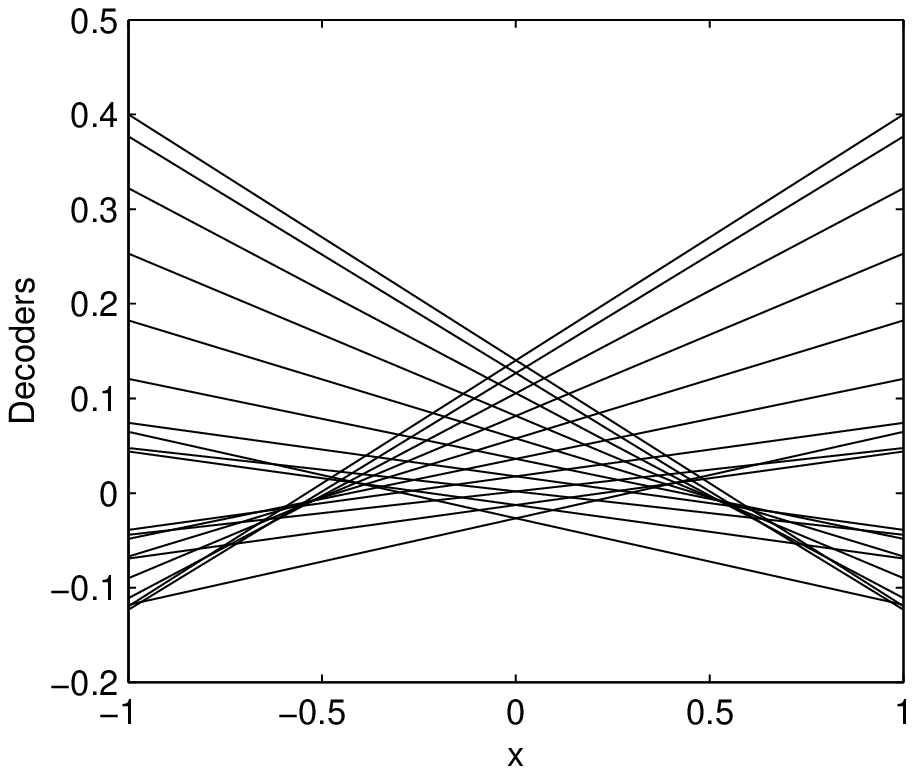}  \\
	\end{tabular}
    \end{center}
    \caption{
    Encoders and decoders obtained using the optimization procedures 
    of sections~\ref{sec:learnencoders} and~\ref{sec:dealwithnoise}.
    (a) 
    An orthonormal basis for the minimal space underlying the 
    piecewise-planar firing rates.  Any line 
    segment over the interval $[-1, 1]$ can be expressed as a linear 
    combination of these two basis functions.  (b) Encoders found using the 
    optimization procedure.  For each neuron, the encoder has a slope  
    identical to that of the firing-rate profile.
    (c) 
    Decoders obtained in the absence of noise require 
    neurons of extreme precision to operate properly.  Two decoders 
    significantly contribute to the decoded PDFs, while all others 
    are zero, contributing  nothing to the decoded PDF.  
    (d)
    Decoders obtained assuming a small amount of noise depend upon 
    all of  the neurons.  The decoders shown here result from a noise variance 
    of 0.01, limiting the precision of the neurons to 
    biologically plausible levels.  
    Note that these decoders can take on 
    negative values, so the functions reconstructed from the neural 
    firing rates may only approximate 
    the encoded PDF.}
    \label{fig:pwlinbasisandencoders}
    \label{fig:pwlindeczerosandnoisy}
\end{figure}

To verify the efficacy of this optimization procedure, we apply it to a set of 
broadly tuned, 
biologically reasonable neuronal responses to a precise input signal.  
In 
particular, we use piecewise-linear activities 
(Figure~\ref{fig:broadroughact}), essentially  one-dimensional versions 
of the response functions entering Georgopoulos's population vector, 
to define our neural responses over the interval $[-1, 1]$
(see also Figure 4 in Fuchs \etal, 1988\nocite{fuchs/etal:1988}).  
We assume a
minimal space spanned by two straight-line functions, shown in 
Figure~\ref{fig:pwlinbasisandencoders}a,
and take the activation function to be 
rectification
\begin{equation}
    f(x) = \left \{
        \begin{array}{cc}
        x & x \geq 0 \\
        0 & x < 0 
    \end{array} \right.
    \label{eq:definerectify}
\end{equation}
  Since we are interested in representing 
a precise input, we choose $\rho(x; t) = 
\delta(x - \xi(t))$.  Applying the optimization procedure, we obtain a set of 
encoders 
(Figure~\ref{fig:pwlinbasisandencoders}b) 
that are able to exactly reconstruct the 
 neural activity patterns with  input 
PDFs of the assumed Dirac delta function form.


\subsection{Obtaining the Decoding Functions}\label{sec:learndecnonoise}\label{sec:dealwithnoise}

A similar procedure is used to find the decoding functions.  
We first  
account for the 
limited precision of neural firing rates and for any intrinsic noise of real 
neurons by converting the neural firing rates into stochastic processes
\begin{equation}
a_{i}(\boldxi)\rightarrow a_{i}(\boldxi) + \varepsilon_{i}
\end{equation}
where $\varepsilon_{i}$ represents the noise source.  We assume \( 
\varepsilon_{i} \) to 
have zero mean without loss of generality; a non-zero mean can be absorbed into 
the firing rate profiles, if needed.  The 
above encoding functions are unchanged by the presence of zero-mean 
noise.

To ensure that the encoders and decoders found are not dependent on a 
particular realization of the noise, we define the cost function
\begin{equation}
    E_{2}  =  {1 \over 2} \left \langle \doubleint \left (\rho(x; \boldxi) - 
    \sum_{i=1}^{N} \left(a_i(\boldxi) + \varepsilon_{i}\right)\phi_i(x)\right)^2 
    \rho(\boldxi) dx d\boldxi \right \rangle_{\set{\varepsilon_{i}}}
\end{equation}
Here, the angle brackets indicate an ensemble 
average over realizations of the neuronal noise.
Substituting equation~\ref{eq:Phitophi} into \( E_{2} \), we 
have
\begin{equation}
    E_{2}  =  {1 \over 2} \left \langle \doubleint \left (\rho(x; \boldxi) - 
    \sum_{i,\nu}
    \Phi_{\nu}(x)\kappa_{\nu i}
    \left(\vphantom{A_{i}^{i}} a_i(\boldxi) + \varepsilon_{i}\right) \right)^2 
    \rho(\boldxi) dx d\boldxi \right \rangle_{\set{\varepsilon_{i}}}
\end{equation}
To find the $\kappa_{\nu i}$ that minimize this cost function, we calculate
\( {\partial E_{2}}/{\partial\kappa_{\nu j}} \).
Taking each $\varepsilon_{i}$ to be independent,
identically distributed, zero-mean 
Gaussian noise with variance $\sigma^{2}$ produces
\begin{equation}
    \frac{\partial E_{2}}{\partial\kappa_{\nu j}} = - M_{\nu j} + 
    \sum_{i}\kappa_{\nu i}\left( \Gamma_{ij} + \sigma^{2} \right )
    \label{eq:definenoisydecoders}
\end{equation}
where
\begin{equation}
    M_{\nu j} = \doubleint \rho(x; \boldxi) a_{j}(\boldxi) \Phi_{\nu}(x) 
    \rho(\boldxi) dx d\boldxi \label{eq:defdecmat}
\end{equation}
and
\begin{equation}
    \Gamma_{ij} = \int a_{i}(\boldxi)a_{j}(\boldxi)\rho(\boldxi)d\boldxi
    \label{eq:defdecgamma}
\end{equation}
Setting the derivatives to zero and recasting 
equation~\ref{eq:defdecgamma} in matrix form, we have
\begin{equation}
\left (\Gamma + \sigma^{2}I \right ) \kappa = M
\end{equation}
We can solve directly for $\kappa$ by inverting $\left ( \Gamma + 
\sigma^{2}I \right )$.

The inclusion of noise is essential for producing sensible decoders.  
To illustrate this fact, we 
determine decoders for neurons with 
piecewise-linear activity patterns employing the basis shown in 
Figure~\ref{fig:pwlinbasisandencoders}a (discussed in 
section~\ref{sec:learnencoders}).
The decoders are used to attempt a reconstruction of the original 
delta-function PDFs, inverting the encoding process previously 
considered.
With \( \sigma^{2} = 0 \), the
algorithm produces two decoders that 
play a significant role 
while the others are all zero 
(Figure~\ref{fig:pwlindeczerosandnoisy}c).  This noise-free solution 
evidently requires neurons that 
are extremely precise in their firing rates, rather than making use of 
redundant neurons to improve the quality of the representation.  
With noise present (\( \sigma^{2} > 0 \)), 
we determine a set of 
decoders that utilizes all of the neurons in the representation 
(Figure~\ref{fig:pwlindeczerosandnoisy}d) 
and 
is independent of 
unrealistically precise firing rates.

\begin{figure}[tbp]
    \centering
    \includegraphics{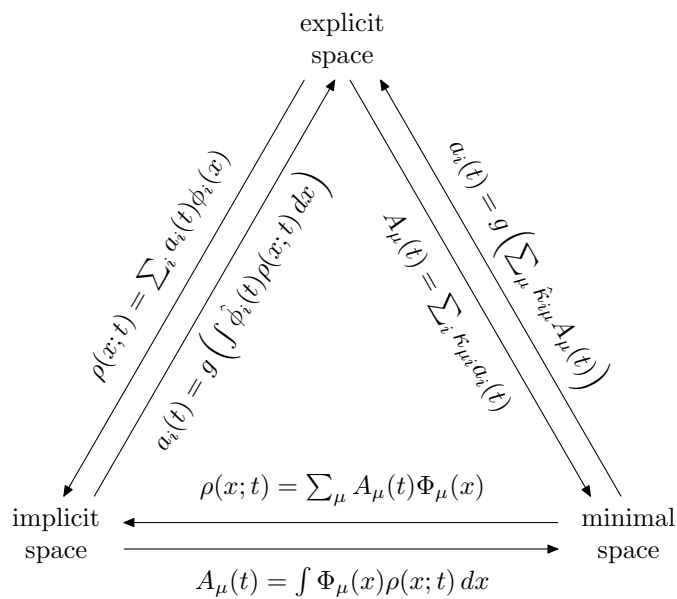}
    \caption{Transformations between the representations.  
    With the indicated rules, we can readily switch 
    between the implicit, explicit, and minimal spaces, associated 
    respectively with the variables \( x \), \( a_{i}\) (\( i = 
    1,2,3,\ldots,N \)), and \( A_{\mu} \) (\( \mu = 1,2,3,\ldots,D 
    \)), and select  
    the most convenient one for any given task. }
    \label{fig:xformrules}
\end{figure}

Having determined the decoders, we can directly 
transform between the explicit, implicit, and minimal spaces.  
The transformation rules are summarized pictorially in 
Figure~\ref{fig:xformrules}.

\subsection{Dimensionality of the Minimal 
Space}\label{subsec:dimminspace}

The structure of the neural representations created depends critically upon the 
dimensionality~\( D \) of the associated spaces.  We can most easily 
explore the effect of  
the dimensionality in the minimal space, where \( D \) is simply equal to 
the number of basis functions~\( \Phi_{\mu}(x) \).

By way of illustration, let us pattern the basis functions after the Legendre polynomials~\( P_{\mu}(x) \).  
The Legendre polynomials form an orthogonal set, but are not 
normalized, so we define
\begin{math}
    \Ptilde_{\mu}(x) = {P_{\mu}(x) }/\sqrt{{\int_{-1}^{1} P_{\mu}^{2}(x) 
    dx}}
\end{math} over the interval~\( [-1,1] \).
For dimension~\( D \), we then set the minimal-space basis 
function \( \Phi_{\mu}(x) \) equal to the  normalized 
Legendre polynomial~\( 
\Ptilde_{\mu-1}(x) \) for \( \mu = 1,2,\ldots D \).

To demonstrate the effect of the dimension~\( D \) upon the quality 
of the neural representation, we compare an assumed target PDF with the 
PDF as represented in neural populations.  We vary \( D \) and generate, 
as described in 
sections~\ref{sec:learnencoders} and~\ref{sec:learndecnonoise}, 
encoding and decoding functions optimized to work with neurons with 
firing rate profiles as shown in Figure~\ref{fig:broadroughact}.  Using 
equation~\ref{eq:encoderequation}, the 
target PDF is encoded into  neural firing rates which are then 
decoded using equation~\ref{eq:decoderequation}.

\begin{figure}[tbp]
    \centering
    \includegraphics[width=3.5in]{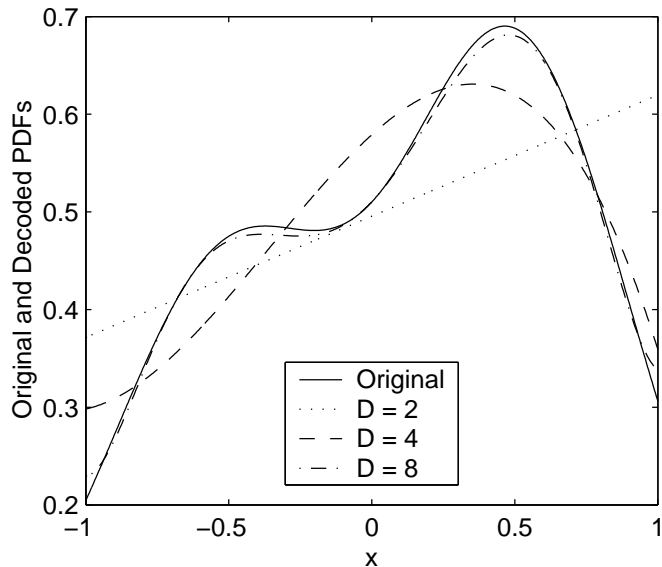}
    \caption{The effect of the dimensionality~\( D \) on the quality 
    of the neural representation.  As~\( D \) is increased, the 
    decoded PDF more closely matches the original PDF.  }
    \label{fig:dimscaling}
\end{figure}

With a bimodal target PDF, increasing \( D \) improves the quality of 
the decoded PDF (Figure~\ref{fig:dimscaling}).  For \( D =2\), only a 
straight line is decoded (although this may still be useful---see 
sections~\ref{sec:neuralintegrator} and~\ref{sec:commchanbroadtuning}), 
while for \( D=8 \), the decoded PDF matches the target PDF quite 
well.  

\subsection{A Neural Integrator Model}\label{sec:neuralintegrator}

An important example of 
a neural integrator is the group of neurons that maintain the eyes 
in a fixed position in the absence of visual input.  These 
recurrently connected neurons are able to hold the eye in position for 
times much longer than the interspike interval of the neurons. 
Collectively, they form an attractor network that acts as a memory of eye 
position which lasts for several seconds \cite{seung:1996}.

By introducing temporal dynamics into the underlying probabilistic 
models, we can create a model of a neural integrator.  The 
dynamics are straightforward: for a short time $\tau$, the PDF should 
be unchanged, so
\begin{equation}
    \rho(x;t+\tau) = \rho(x;t)
    \label{eq:neurintprobmod}
\end{equation}
where $x$ is the value (i.e.\ eye position) stored in the memory.  

As discussed above, we generate decoding functions using 
piecewise-linear activities, linear encoders, and a rectifying 
activation function.
Making use of this representation, the encoding and decoding rules 
(equations~\ref{eq:decoderequation} and~\ref{eq:encoderequation}),
and the probabilistic dynamics 
(equation~\ref{eq:neurintprobmod}), we can show that
\begin{eqnarray}
    a_{i}(t+\tau) & = & g \left(\int \phihat_{i}(x) \rho(x;t+\tau) dx\right) \\
    & = & g \left(\sum_{j}a_{j}(t) \int \phihat_{i}(x) \phi_{j}(x) dx\right) 
\end{eqnarray}
Defining weights
\begin{equation}
    \omega_{ij} = \int \phihat_{i}(x) \phi_{j}(x) dx
    \label{eq:neurintwts}
\end{equation}
we may rewrite this as
\begin{equation}
    a_{i}(t+\tau) = g \left ( \sum_{j}\omega_{ij}a_{j}(t) \right)
    \label{eq:neurinteq}
\end{equation}
The recurrent neural network that results is fully connected, with each 
neuron having a synaptic connection to every other neuron.

\begin{figure}[tbp]
    \centering
    \includegraphics[width=3.5in]{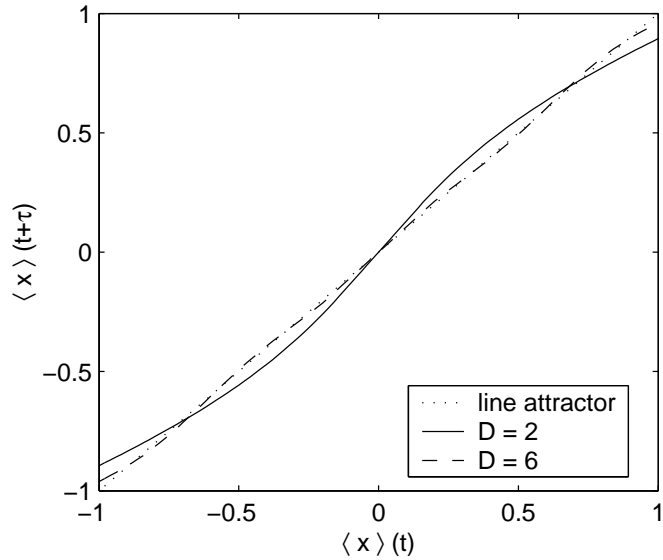}
    \caption{The neural integrator model maintains only a limited number of 
    values, rather than an arbitrary input value.  The number of stable fixed 
    points of the neural integrator model can be seen in the network's transfer 
    function.  Here there are two stable fixed points for a 
    neural integrator consisting of 20~neurons with encoders and decoders 
    found using 
    a minimal space with dimension \( D = 2 \).  By increasing \( D  \) 
    to~4, the number of stable fixed points increases to~3 
    (not shown), while increasing \( D \) to~6 yields~4 stable 
    fixed points.  With only the 20~neurons of limited precision  
    utilized here, further increases in \( D \) do not give rise to 
    further increases in the number of stable fixed points.    }
    \label{fig:neurintxferfixpt}
\end{figure}

The stored value of the eye position is extracted by calculating the
expectation value of the random variable $x$, weighted by the decoded 
PDF.  Ideally, we would like any value in the supported range to be 
held constant, so that the network functions as a line attractor 
\cite{seung:1996}, a kind of continuous attractor.  However, the system actually
operates as a
collection of point attractors with only a limited number of stable 
fixed points, as can be seen from the network's transfer function 
(Figure~\ref{fig:neurintxferfixpt}).  
The structure of the transfer function, and the number of stable fixed 
points, depends on the dimensionality \(D\) of the minimal space.
As the dimensionality of the  minimal space is increased, the 
neural integrator can support 
additional stable fixed points, eventually approximating a line attractor. 
This neural integrator model is essentially a variation of the model 
constructed
by Eliasmith and Anderson (1999\nocite{eliasmith/anderson:1999}).

\section{Probabilistic Inference Performed by Neural Networks}

\subsection{Inference}
Inference between two related variables \( x \) and \( y \)
in the implicit space
is performed by taking a weighted 
average of the conditional probability $\rho(y\given x)$:
\begin{equation}
\rho(y;t+\tau) = \int \rho(y\given x) \rho(x;t) dx
\label{eq:justifyinference}
\end{equation}
We have assumed in equation~\ref{eq:justifyinference} that the 
relationship between $x$ and $y$ is independent of the values of the 
minimal parameters, so
\begin{equation}
	\rho(y\given x; \set{A_{\mu}(t)}) = \rho(y \given x)
\end{equation}
This assumption fixes the structure of the probabilistic model, explicitly 
excluding learning from any neural networks derived from it.  The 
conditional probability $\rho(y\given x)$ is like a fixed 
look-up-table; the Marr-Albus theory of cerebellar function can be 
directly mapped into equation~\ref{eq:justifyinference} 
\cite{hakimian/etal:1999}.

Mapping the implicit-space 
inference relation~\ref{eq:justifyinference} into the explicit space 
of neurons yields a neural network (Anderson, 1994, 1996; Zemel and 
Dayan, 1997)\nocite{anderson:1994,zemel/dayan:1997}.
Specifically, one imposes representations as given in equations~\ref{eq:basisdec} 
and~\ref{eq:xencoderule} for \( x \), and 
\begin{eqnarray}
	\label{eq:ydecoderule}
	\rho(y;t) & = & \sum_j b_j(t) \psi_j(y)\\
	b_j(t) & = & g\left(\int \psihat_j(y) \rho(y;t) dy \right)
\end{eqnarray}
for \( y \).  Then one combines these representations with 
equation~\ref{eq:justifyinference},  
leading to
\begin{equation}
	b_j(t+\tau) = g\left(\sum_i w_{ji} a_i(t)\right)
	\label{eq:ffeq}
\end{equation}
with the coupling coefficients
\begin{equation}
w_{ji} = \doubleint \psihat_j(y) \rho(y\given x) \phi_i(x) dx dy
	\label{eq:ffwts}
\end{equation}
For well-chosen encoding and decoding functions, 
equations~\ref{eq:ffeq} and~\ref{eq:ffwts} allow us to construct a neural 
network that embodies the desired relationship between the implicit variables, 
without applying a training procedure to find a relation from a data set.

This approach to inference is naturally extended to greater numbers of 
implicit variables.  For example, suppose we add a second input~$z$ to 
the above network, and write
\begin{equation}
\rho(y;t+\tau) = \doubleint \rho(y\given x, z) \rho(x;t) \rho(z;t) dx dz
\end{equation}
Representing $z$ using
\begin{eqnarray}
	\rho(z;t) & = & \sum_k c_k(t) \theta_k(z)\\
	c_k(t) & = & f\left(\int \thetahat_k(z) \rho(z;t) dx\right)
\end{eqnarray}
leads to 
\begin{equation}
b_j(t+\tau) = g\left(\sum_i w_{jik} a_i(t) c_k(t)\right)
\end{equation}
with
\begin{equation}
w_{jik} = \tripleint \psihat_j(y) \rho(y\given x,z) \phi_i(x) \theta_k(z) dx 
dy dz
\end{equation}
An interesting feature of this neural network is that it employs
multiplicative interactions.  This multiplication might be realized
by coincidence detection in the dendrites; the implication is 
that the dendrites are active processing elements 
\cite{mel:1994,cash/yuste:1998}.

\subsection{A Communication Channel Model}\label{sec:commchanbroadtuning}

As a concrete example of probabilistic inference within the PDF scheme,
we now use equations~\ref{eq:ffeq} 
and~\ref{eq:ffwts} to implement a 
communication channel.  Specifically, we wish to encode a single input 
value~$\xi(t)$ into a PDF $\rho(x; t)$ represented by a population of 
neurons, and copy that PDF into another PDF $\rho(y; t)$ represented 
by a second population of neurons.  To extract a unique output value 
from $\rho(y;t)$, we focus on the expectation value of $y$.  
We use 20~neurons to represent the input PDF $\rho(x;t)$ and 
16~neurons to represent the output PDF $\rho(y;t)$.  The encoders and 
decoders for these neurons are generated from two straight-line basis 
functions (Figure~\ref{fig:pwlinbasisandencoders}a) and piecewise-linear neural 
responses as explained previously (sections~\ref{sec:learnencoders} 
and~\ref{sec:dealwithnoise}).

Since we only want to encode a single value, and not a complex 
multimodal distribution, we describe the input using a PDF of the form 
$\rho(x;t) = \delta(x-\xi(t))$.  We set the form of the conditional PDF to 
be $\rho(y \given x) = \delta(y - x)$; accordingly, in the implicit space, we 
expect that \( \rho(y;t) = \delta(y - \xi(t)) \).  However, a PDF with such a 
delta-function form is quite intractable in the explicit space---no 
finite linear combination of functions can yield the expected form of 
\( \rho(y;t) \). 
Our goal  is thus to obtain an accurate estimate of \( \xi(t) \), 
rather than a perfect reconstruction of the PDFs.

\begin{figure}[tbp]
    \centering
    \includegraphics{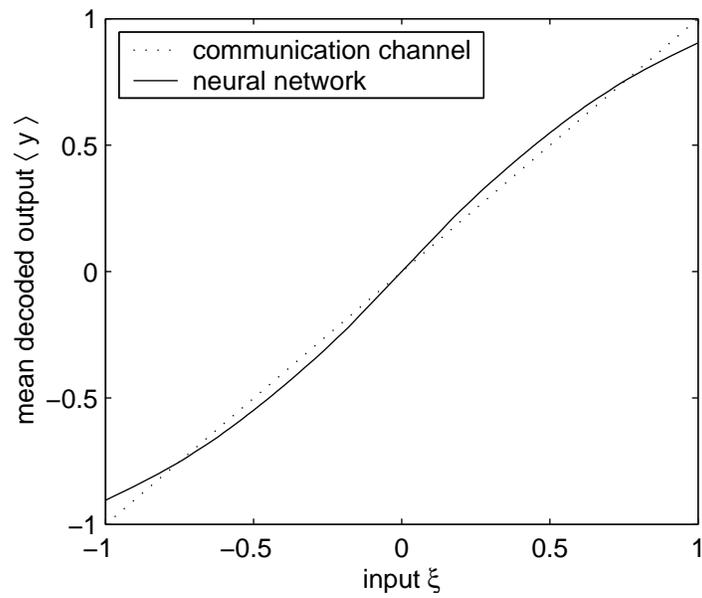}
    \caption{Although the PDF is not accurately represented, the mean 
    value of the PDF can be satisfactorily retrieved from the neural network.  
    By using only two basis functions to generate the decoders, the 
    output PDFs are elements of a space of dimension two.  This is 
    suitable for representing the total weight and the mean of a PDF, 
    but not higher moments.}
    \label{fig:pwlinxferfnctn}
\end{figure}

To interpret the performance of the neural network, we compare the 
expectation value $\langle y \rangle$  
(weighted by the PDF decoded from the network outputs \set{b_{j}(t)}) 
to the input 
$\xi$.  The decoded PDF
is a weighted sum of linear decoding functions, and is thus a straight 
line itself.  This is of course a poor reproduction of the Dirac 
delta function input,
but $\langle y \rangle$ is closely 
in accord with the input values (fig~\ref{fig:pwlinxferfnctn}). 
We 
may understand this by considering the basis functions used:  they are 
well-suited for calculating the 0th and 1st moments of the PDF, but 
unsuitable for calculating higher-order moments.

\subsection{Working in the Minimal Space}

So far, we have used the concept of the minimal space as a tool for 
developing the encoders and decoders.  However, we also can make 
direct use of the minimal space to set up abstract networks, then 
convert 
those into networks of real neurons.  To accomplish this, we 
derive 
relations between the firing rates in the two spaces (\set{A_{\mu}(t)} 
and \set{a_{i}(t)}).
The neural network in the explicit space then constitutes a physical implementation 
of the abstract network in the minimal space.  The issues of the 
role of neuronal firing rate variability  in the population code (see for example Abbott and 
Dayan, 1999\nocite{abbott/dayan:1999}) may thus be separated from the issues of the propagation 
of probabilistic information.

First, consider the decoding rules given by 
equations~\ref{eq:basisdec} and~\ref{eq:decoderequation}.
Making use of equation~\ref{eq:Phitophi}, we obtain
\begin{equation}
\sum_{\mu}A_{\mu}(t)\Phi_{\mu}(x) = \sum_{i}a_{i}(t)\sum_{\nu} 
\kappa_{\nu i}\Phi_{\nu}(x)
\end{equation}
Since the $\Phi_{\mu}(x)$ are orthonormal functions, we have
\begin{equation}
A_{\mu}(t) = \sum_{i}\kappa_{\nu i}a_{i}(t) \label{eq:ntomn}
\end{equation}
for  transforming from the explicit space to the minimal space.

Next, consider the encoding rule given by equation~\ref{eq:encoderequation}.
Recalling that $\phihat_{i}(x) = 
\sum_{\nu}\kappahat_{i\nu}\Phi_{\nu}(x)$ and $A_{\nu}(t) = \int 
\Phi_{\nu}(x) \rho(x; t) dx$, we have
\begin{equation}
    a_{i}(t) = f \left ( \sum_{\nu}\kappahat_{i\nu} 
    A_{\nu}(t) \right ) \label{eq:mnton}
\end{equation}
for transforming from the minimal space to the explicit space.

Using equations~\ref{eq:ntomn} and \ref{eq:mnton}, we can translate 
between the minimal and explicit spaces.  This allows us to set up 
neural networks by first working in the mathematically convenient 
minimal space.   To illustrate this procedure, 
we return
to the $X \longrightarrow Y$ 
inference network. 
We take the minimal spaces for both 
the input $x$ and the output $y$ to be defined by linear functions 
over the interval~$[-1, 1]$, with 
basis functions 
of the form shown in 
Figure~\ref{fig:pwlinbasisandencoders}a.  The associated PDFs are represented 
using equation~\ref{eq:basisdec} and 
\begin{equation}
	\rho(y;t) = \sum_{\nu} B_{\nu}(t) \Psi_{\nu}(y)
\end{equation}
With these representations, the probabilistic relation given in 
equation~\ref{eq:justifyinference}
becomes
\begin{equation}
B_{\nu}(t+\tau) = \sum_{\mu}\Omega_{\nu \mu} A_{\mu}(t) \label{eq:underspacenet}
\end{equation}
where
\begin{equation}
	\Omega_{\nu \mu} = \int \Psi_{\nu}(y) \rho(y \given x) \Phi_{\mu}(x) dx dy
\end{equation}
We next convert this into a neural network in the explicit space 
using equations~\ref{eq:ntomn} and~\ref{eq:mnton},
so that
\begin{eqnarray}
b_{j}(t+\tau) & = & g \left (\sum_{\nu}\kappahat^{(y)}_{j\nu}B_{\nu}(t) 
\right ) \nonumber \\
& = & g\left(\sum_{i} \sum_{\mu, \nu}\kappahat^{(y)}_{j\nu}\Omega_{\nu \mu} 
\kappa^{(x)}_{\mu i}a_{i}(t) \right)
\label{eq:undertoexplicit}
\end{eqnarray}
By identifying
\begin{equation}
\omega_{ji} = \sum_{\mu, \nu}\kappahat^{(y)}_{j\nu}\Omega_{\nu \mu} 
\kappa^{(x)}_{\mu i}
\end{equation}
we may rewrite equation~\ref{eq:undertoexplicit} as
\begin{equation}
b_{j}(t+\tau) = g \left ( \sum_{i} \omega_{ji} a_{i}(t) \right )
\end{equation}
arriving at a neural network with the same feedforward 
dynamics (equation~\ref{eq:ffeq}) and the same synaptic weights
(equation~\ref{eq:ffwts}) found 
previously.  

This example  reproduces results previously found by working in the 
explicit space, but also highlights several advantages of working in
the minimal space.  
Perhaps most importantly, the fundamental structure of the neural
networks is made more transparent by eliminating the
redundancies that arise in the networks due to the limited 
representational ability of neurons.  Significantly, we see that 
computational properties of the nonlinear update rule for the output neurons 
(equation~\ref{eq:undertoexplicit}) can be understood by studying the 
linear update rule in the minimal space 
(equation~\ref{eq:underspacenet}),
consistent with the population vector representations investigated by 
Georgopoulos \etal\ (1986\nocite{georgopoulos/etal:1986}).

\section{Conclusions}

We have examined some of the ramifications of the hypothesis that neural 
networks represent information as probability density functions.  
These PDFs are assumed to be expressible a linear combination of some 
implicit decoding functions, with the decoder for each neuron being weighted 
by its firing rate.  The firing rates in turn may be obtained from 
a PDF using a complementary set of encoding functions.

In general, the encoding and decoding functions that we have introduced 
are numerous enough to define
spaces of very high dimension, far beyond the range of accurate
representation by biological neurons having a precision 
of only a few bits.
To mediate this conflict between computational requirements and
biological reality, we have introduced an auxiliary representation of
a lower-dimensional minimal space appropriate to the nature and scale
of the computations that neurobiological systems actually perform
on the relevant input and output analog variables.
The basis functions in this minimal space are 
used to represent both the encoding and decoding functions, limiting 
the dimensionality of the spaces they define.  As an added benefit, 
the minimal space---and the associated metaneuron variables---can be 
chosen to have properties that facilitate
theoretical characterization of the neural networks 
resulting from the PDF hypothesis.

%
These neural networks  are based upon the available
probabilistic 
information and upon the encoding and decoding functions.  The synaptic  
weights of the networks are fully specified without a training 
procedure.  A natural extension of the work we have presented is the 
addition of learning rules for determining the weights.  Learning rules would 
provide several advantages; in particular, 
they would facilitate the generation of neural networks when  data is 
available but the 
underlying computations are not entirely clear.  The optimization 
procedure we utilized to find the encoding functions may be a useful 
starting point for identifying a more complete learning rule. 

Researchers in the fields of molecular biology, immunology, genetics, 
development, and evolution, all of which involve highly complex systems having
many degrees of freedom, are beginning to explore the use of ``metavariables''
as a formal means to reduce the dimensionality of the space of parameters
that must be dealt with in achieving viable and tractable quantitative
descriptions.  The formal results we have derived for metavariable
(``metaneuron'') representation of function spaces and the experience 
we have gained through associated model simulations may prove valuable 
for parallel investigations in these 
and other
fields.

Returning to the neurobiological context, we may comment on the
the role that is envisioned for the PDF formalism in the modeling of 
brain function.  Recent work based on population-temporal coding 
(e.g. Eliasmith and Anderson 1999, 
2002\nocite{eliasmith/anderson:1999,eliasmith/anderson:2002}) 
indicates that the modeling of low-level 
sensory processing and output motor control do not require such 
a sophisticated representation; manipulation of mean values is 
generally sufficient and the representations can be simplified 
to deal with vector spaces instead of function spaces.  However,
explicit representation of probabilistic descriptors of the state 
of knowledge of pertinent analog variables may prove indispensible
to an understanding of higher-level processes.  For example, estimates 
of depth at each spatial location from the disparity between the 
images impinging on both eyes can never be made with precision using 
a purely bottom-up strategy. 

The modern approach to all higher-level image-processing tasks is driven by  
the theory of Bayesian inference, in which models are developed and 
parameters estimated based on a set of well-defined rules within 
a probabilistic framework.  In a second paper 
\cite{barber/clark/anderson:2001b}, we 
carry the PDF program a step further by formulating procedures for
embedding joint probabilities into neural networks.  These procedures
allow us to design neural circuit models that pool multiple sources 
of evidence.  In our view, this offers the most rational approach 
to building and understanding cortical circuits that carry out well-posed 
information-processing tasks.

\section*{Acknowledgements}

This research was supported by the National Science Foundation under
Grant Nos. IBN-9634314, PHY-9602127, and PHY-9900713.  
While writing this work, MJB was a member of the Graduiertenkolleg 
Azentrische Kristalle, GK549 of the DFG; the writing was completed 
while MJB and JWC were participants 
in the Research Year on ``The Sciences of Complexity: From Mathematics 
to Technology to a Sustainable World'' at the Zentrum f\"ur 
interdisziplin\"are Forschung, University of Bielefeld.

\bibliographystyle{apalike}
\bibliography{neurcomp,mjbpubs}

\end{document}